\documentclass[11pt,a4paper]{article}%
\usepackage[T1]{fontenc}
\usepackage[latin1,utf8]{inputenc}

\pdfoutput=1

\usepackage{amsmath,amssymb,amsfonts,multicol}
\usepackage[normal,font=small,labelfont=bf,labelsep=period]{caption}
\usepackage[pdftex]{color,graphicx}
\usepackage[english]{babel}
\usepackage[compress]{cite}

\usepackage[dvipsnames]{xcolor}
\definecolor{darkgreen}{rgb}{0,0.65,0}
\usepackage{slashed}

\usepackage{ulem}

\usepackage[linktoc=all,colorlinks=true,linkcolor=black,urlcolor=magenta,citecolor=black]{hyperref}
\hypersetup{%
  colorlinks = true,
  linkcolor  = black
}
\newcommand{\be}{\begin{equation}}
\newcommand{\ee}{\end{equation}}
\newcommand{\bea}{\begin{eqnarray}}
\newcommand{\eea}{\end{eqnarray}}

\newcommand{\eq}[1]{(\ref{#1})}

\definecolor{gb}{rgb}{0.8,0.3,0.1}

\definecolor{rev}{rgb}{1,0,0}

\numberwithin{equation}{section}

\begin{document}
\begin{flushright}
\footnotesize
{IFT-UAM/CSIC-18-74}\\
{LAPTH-027/18}

\end{flushright}
\color{black}

\begin{center}
{\LARGE\color{black}\bf On the merger rate of primordial black holes:\\ effects of nearest neighbours distribution and clustering\\[1mm] }

\medskip
\bigskip\color{black}\vspace{0.6cm}

{
{\large\bf Guillermo Ballesteros,}\ $^{a}$
{\large\bf Pasquale D. Serpico,}\ $^{b}$
{\large\bf Marco Taoso}\ $^{c}$
}
\\[7mm]

{\it $^a$  Instituto de F\'isica Te\'orica UAM/CSIC,\\Calle Nicol\'as Cabrera 13-15, Cantoblanco E-28049 Madrid, Spain}\\[3mm]
{\it $^b$  Univ. Grenoble Alpes, USMB, CNRS, LAPTh, F-74940 Annecy, France}\\[3mm]
{\it $^c$  INFN, Sez. di Torino, via P. Giuria, 1, I-10125 Torino, Italy}\\[3mm]
\end{center}

\bigskip

\centerline{\large\bf Abstract}
\begin{quote}
\color{black}\large 
One of the seemingly strongest constraints on the fraction of dark matter in the form of  primordial black holes (PBH) of ${\cal O}$(10)$\,M_\odot$ relies on the
merger rate inferred from the binary BH merger events detected by LIGO/Virgo. The robustness of these bounds depends however on the accuracy with which the formation of PBH binaries
in the early Universe can be described.  We revisit the standard estimate of the merger rate, focusing on a couple of key ingredients: the 
spatial distribution of nearest neighbours and the initial clustering of PBHs associated to a given primordial power spectrum. 
Overall, we confirm the robustness of the results presented in the literature in the case of a narrow mass function (which constrain the PBH fraction of dark matter to be $f_{\rm PBH}\lesssim 0.001-0.01$).
The initial clustering of PBHs might have an effect tightening the current constraint, but only for very broad mass functions, corresponding to wide
bumps in the primordial power spectra extending at least over a couple of decades in $k$-space.
\end{quote}

\begin{center} 

\vfill\flushleft
\noindent\rule{6cm}{0.4pt}\\
{\small  \tt guillermo.ballesteros@uam.es, serpico@lapth.cnrs.fr, marco.taoso@to.infn.it }

\end{center}

\newpage

\section{Introduction}
\label{sec:introduction}

The recent LIGO/Virgo discoveries of binary black hole (BH) and neutron star (NS) mergers have  ushered us in the era of gravitational wave (GW) astronomy~\cite{Abbott:2016blz,Abbott:2016nmj,Abbott:2017vtc,TheLIGOScientific:2017qsa}, leading in a short time to interesting implications on several branches of physics. In particular, even though only a single NS-NS merger has been detected so far~\cite{TheLIGOScientific:2017qsa}, its multi-messenger observation \cite{GBM:2017lvd} has permitted to draw a number of inferences, ranging from constraints on the nuclear equation of state~\cite{De:2018uhw,Abbott:2018exr} and some alternative
models of gravity at large scales \cite{Creminelli:2017sry,Sakstein:2017xjx,Ezquiaga:2017ekz,Baker:2017hug} (a possibility anticipated in~\cite{Lombriser:2015sxa})
to the identification of an 
engine of short gamma-ray bursts~\cite{Goldstein:2017mmi} 
and $r$-process (and thus the nucleosynthesis of heavy elements)~\cite{Drout:2017ijr}.
  
On the other hand,  the origin of the progenitors of the few observed binary BH mergers has yet to be determined. The high value of some of their mean component masses (above 30 $M_\odot$) are suggestive of progenitor formation in low-metallicity environments; see e.g.~\cite{LOwM}. Besides, classical isolated binary evolution via mass transfer including a common envelope phase has
been shown to be capable of reproducing the first detected events with a single channel~\cite{Stevenson:2017tfq}. Yet,  many uncertainties remain in
the assumptions and parameters of astrophysical models of BH (and binary BH) formation; for a compact review, see~\cite{Mapelli:2018baq}. For instance, a dynamical origin of the binaries, notably in young massive and open stellar clusters, is a viable (albeit uncertain) possibility, see e.g.~\cite{Banerjee:2016ths,Chatterjee:2016hxc}. Furthermore, a multi-channel origin,  in particular the idea that some or all of the mergers could be primordial --a proposal dating back to \cite{Carr:1974nx}; see~\cite{Sasaki:2018dmp} for a recent review-- cannot be disregarded at the moment~\cite{Mapelli:2018baq}.

Soon after the first merger was detected, some authors \cite{Bird:2016dcv,Clesse:2016vqa} went as far as conjecturing
that the population of BHs responsible for these mergers is, in fact, not only primordial, but also accounts for the still unidentified  dark matter (DM) which 
constitutes about one quarter of the energy density of the Universe \cite{Ade:2015xua}. 
The
plausibility of this conjecture is, however, very much debated. 
Any mechanism able to produce primordial black holes (PBHs) in the {\it stellar} ($\sim 1$ -- $100\, M_\odot$) mass range has to 
comply with constraints coming from a number of considerations:\footnote{Only the most recent and constraining analyses of each kind known to us are reported in this list.} 
the  half-light radius~\cite{Brandt:2016aco,Li:2016utv} or the stellar distribution~\cite{Koushiappas:2017chw} of dwarf galaxies, Galactic microlensing~\cite{Tisserand:2006zx,Wyrzykowski:2011tr,Green:2016xgy}, quasar microlensing~\cite{Mediavilla:2017bok}, lensing of type-Ia supernovae~\cite{Zumalacarregui:2017qqd}, the orbital dynamics of halo wide binaries~\cite{Monroy-Rodriguez:2014ula}, radio and X-ray observations of the Milky Way~\cite{Gaggero:2016dpq,Hektor:2018rul}, pulsar timing~\cite{Schutz:2016khr}, and, importantly since probing a pristine phenomenon, the Cosmic Microwave Background (CMB)~\cite{Ali-Haimoud:2016mbv,Poulin:2017bwe}.  A weaker but independent constraint on the viability of PBH DM 
comes from the non-observation of a stochastic GW background (due to the mergers of  PBH binaries at high-$z$, in the matter dominated era) during the first aLIGO run~\cite{Wang:2016ana}; see also~\cite{Clesse:2016ajp}. Finally, another such background (at $\sim$ nHz frequencies) 
could have been produced in association to PBH formation of the appropriate masses in the very early Universe (assuming an inflationary origin of PBHs), as well as spectral ($\mu$-type) distortions of the CMB~\cite{Inomata:2016rbd}.
It is worth noting that the latter bounds are rather model dependent and could be evaded, e.g.\ if the fluctuations leading to PBH formation are strongly non-Gaussian~\cite{Nakama:2016gzw}.

A crucial point --since it concerns  the very self-consistency of the scenario of stellar mass PBH DM-- is the  (or lack of)  compatibility with the inferred merger rate from the LIGO/Virgo detections.
The first recent analyses~\cite{Bird:2016dcv,Clesse:2016vqa} 
underestimated the merger rate of PBHs, by assuming that the binaries were formed in halos, hence in the late Universe.
However, as shown in~\cite{Sasaki:2016jop} and consistently with the first estimates in~\cite{Nakamura:1997sm}, the currently observed binary merger rate of PBHs should actually be dominated (by orders of magnitude) by binaries formed in the early Universe, before matter-radiation equality. The estimated rates from the reported LIGO/Virgo detections range from approximately 1 to 100 events per year and cubic Gpc \cite{Abbott:2017iws,Abbott:2016drs,Abbott:2016nhf}, which limits the PBH contribution to DM in the stellar mass range to sub-percent level. Taken at face value, this bound is the strongest one so far on the PBH abundance for masses $\mathcal{O}(10)M_\odot$.

This conclusion is robust as long as the orbital parameters of the PBH binaries formed in the early Universe are not strongly affected by processes intervening between the times of formation and merger.
Some approximations made in the initial estimate~\cite{Nakamura:1997sm}  were checked already one year later, in~\cite{Ioka:1998nz}, and found to be correct at the level of $\sim 30\%$ each. The aforementioned bound has been recently confirmed in~\cite{Ali-Haimoud:2017rtz},  where its robustness was checked with respect to the effect of tidal fields of halos and interactions with other PBHs,  dynamical friction by ``standard'' DM particles --see also \cite{Kavanagh:2018ggo} for a complementary study of the effect of DM particles-- and, to some extent also, the effect of  baryon accretion.  It has also been claimed that considering a broadly distributed mass function (instead of assuming a very narrow one) does not alter  appreciably the conclusion either~\cite{Raidal:2017mfl,Chen:2018czv}.
  
In order to further  investigate the robustness of this self-consistency test of the idea that DM could be in the form of  stellar mass PBHs, we take a complementary path to the ones explored so far, aiming to check the reliability of the  merger rate calculation with respect to the statistical distribution of the initial orbital parameters. An obvious reason to pose this question is that all previous analyses have considered uniformly or randomly distributed PBHs whereas, in  reality, some degree of clustering can be expected, given that PBHs are supposed to form from the collapse of large density fluctuations. This point has been briefly addressed in~\cite{Raidal:2017mfl} with a parametric study, finding what appear to be potentially large effects. However, the actual impact of PBH clustering (and its dependence on the PBH formation mechanism)
should be assessed in full depth, a task which we start tackling in this article. 

As we will discuss, the formation of early PBH binaries --before the time of matter radiation-equality-- and the subsequent merger rate depend on the  nearest neighbour statistics of PBHs --which determines the orbital semi-axis of the binary-- and the next-to-nearest distribution too, which is responsible for the angular momentum (or, equivalently, the eccentricity) of the binary that forms; see \cite{Nakamura:1997sm,Sasaki:2016jop,Ali-Haimoud:2017rtz,Kavanagh:2018ggo,Raidal:2017mfl} for earlier treatments of the problem.

This article is structured as follows: in Sec.~\ref{basicf}, we summarize the basic formalism and the equations needed for linking the primordial input parameters to the PBH orbital elements and, eventually, the merger time. The reader with a professional knowledge of the literature can simply  skim through this section, which also serves to set our notation. 
In Sec.~\ref{NNdist}, we provide a simple derivation of 
the probability distribution function (PDF) of PBH inter-distances in the early Universe.
A more formal proof is reported in Appendix~\ref{derivation}.
In Sec.~\ref{correl}, we describe the key aspects of the initial PBH clustering, focusing on its statistics. Finally, in Sec.~\ref{sec:conclusions} we discuss our findings, future perspectives and present our conclusions.

\section{Binary PBH mergers} \label{basicf}

The relevant mechanism for the formation of PBH binaries was proposed more than twenty years ago in \cite{Nakamura:1997sm} (see also \cite{Ioka:1998nz}) and the merger rate estimate has recently been reviewed and refined, see e.g.\ \cite{Sasaki:2016jop,Ali-Haimoud:2017rtz,Sasaki:2018dmp}.  For negligible initial velocities, the basic condition for the formation of a PBH binary is that the gravitational attraction between two PBHs dominates over the separating pull due to the expansion of the Universe. Neglecting all other influences, the two objects would end up in a heads-on collision. In reality, a perturbation with respect to this idealized situation induces an  angular momentum, leading to the formation of a (usually highly eccentric) binary. 
Following arguments similar to the classical treatment of~\cite{Chandrasekhar:1943ws}, it can be shown that the leading role in generating the angular momentum is provided by the nearest PBH to the pair. The gravitational force between this PBH and the other two is subdominant, such that it does not end up forming a bound state with them. Already from these simple considerations, it seems reasonable to expect that the statistical properties of the possible configurations of three PBHs could have a quantitave effect on the merger rate of (early) PBH binaries. However, the original work of \cite{Nakamura:1997sm} as well as most of the subsequent works (e.g. \cite{Sasaki:2016jop,Ali-Haimoud:2017rtz,Sasaki:2018dmp}) assumed that, initially, the PBHs are uniformly distributed in space. Clearly, a more general treatment that includes the statistics of the nearest  and the next-to-nearest neighbours (NN and NtNN, respectively) for any PBH distribution is possible. Such an analysis is also instrumental in including the effect of the PBH clustering, as we will discuss below. 

Let us start by reviewing the basic equations that characterize a PBH binary. It can be easily checked that the gravitational attraction between two approximately isolated PBHs dominates their dynamics if their average mass is above the background mass contained in a comoving sphere whose radius is equal to the separation between them. This can occur during the radiation epoch, due to the different scaling with time of the two competing effects (their mutual interaction versus the pull of expansion) in the equation of motion for their separation \cite{Ali-Haimoud:2017rtz}. More quantitatively, and assuming that all the PBHs are of the same mass, $M$, two near-neighbour PBHs decouple from the Hubble flow during radiation domination provided that their comoving separation, $x$,  approximately satisfies
\begin{equation}
x <  x_{\rm max} \equiv (f_{\rm PBH}/n)^{1/3}\,,
\label{eqn:condition_dec}
\end{equation} 
where $f_{\rm PBH}$ is the fraction of DM in the form of PBHs and the constant $n$ is the average comoving number density of PBHs. 
Their decoupling from the background dynamics occurs at a redshift $z_{\rm dec}$:
\begin{equation}
1+z_{\rm dec}=(1+z_{\rm eq}) \left({x_{\rm max}}/{x}\right)^3\,,
\end{equation}
where $z_{\rm eq}\simeq 3400$ is the redshift of matter-radiation equality and $x$ is the initial comoving separation of the PBHs, assuming here and throughout negligible initial peculiar velocities. 
Therefore, given $x$ and the PBH masses, the decoupling time is determined by the PBH distribution. The latter can be approximated with the abundance at PBH formation, assuming that accretion and evaporation are negligible before the epoch of binary formation.  

The geometry of the initial elliptical orbit is determined by two quantities, which can be chosen as the ellipticity and the semi-major axis of the binary. The latter, which we denote by ${\sf a}$, is determined by $z_{\rm dec}$ and $x$ as follows: 
\begin{equation}\label{semi}
{\sf a}=\frac{x}{1+z_{\rm dec}}=\frac{\Omega_{\rm DM}\,\rho_{c}\,x^4}{(1+z_{\rm eq})M}\,,
\end{equation}
where we have used that $\rho_{\rm PBH}/f_{\rm PBH}=\Omega_{\rm DM}\,\rho_{c}$ is the current DM energy density and $\rho_{c}$ denotes the present critical density of the Universe. The approximation \eq{semi} reproduces well the result obtained numerically in \cite{Ali-Haimoud:2017rtz} by solving the dynamical equation for the separation of the PBHs. Assuming that the angular momentum that induces the orbit is provided by a third PBH at an initial comoving distance $y$ from the center of the binary progenitors, the ellipticity ${\sf e}$ is \begin{equation} \label{ell}
{\sf e}\simeq \sqrt{1-\left({x}/{y}\right)^6}\,,
\end{equation}
where we have neglected an order one factor in front of the ratio $(x/y)^6$.
Clearly, ${\sf e}\simeq 1$ for $y\gg x$, which is the typical situation for standard PBH formation mechanisms, in which each Hubble patch is scarcely populated by PBHs when they form.
Up to a (small) error of the order $x/y$, this also makes $y$ essentially equal to both the distance to the binary center and to each of the binary constituents. 
The binary merges through the emission of gravitational radiation at a time $t_{\rm m}$ after its formation, which can be estimated as~\cite{Peters:1964zz}
\begin{equation} \label{tmerge}
t_{\rm m}\simeq\frac{3 \left(1-{\sf e}^2\right)^{7/2}{\sf a}^4}{170\, G_{\rm N}^3\,M^3}\,,
\end{equation}
where $G_{\rm N}$ denotes Newton's gravitational constant.\footnote{This approximation to the actual merger time should be used with some caution since, for instance, it breaks down at sufficiently early times, when formally the merger time becomes shorter than the free-fall one.}

The only missing ingredient that we need is the spatial distribution of PBHs. We can express the differential probability (per unit time) to form a binary system that will merge at a time $t$ as follows:
\begin{equation}
{\rm d P} = \frac{1}{2}  Q(x,y)\, \theta(y-x)\, \theta(x_{\rm max}-x)\, \delta(t-t_{\rm m}(x,y))\,{\rm d} x\, {\rm d} y\,.
\label{eqn:probability3BH_1}
\end{equation}
The $\delta(t-t_{\rm m}(x,y))$ ensures that the PBHs merge at the time $t_{\rm{ m}}$, and is defined via the eqs.~\eq{semi}, \eq{ell} and \eq{tmerge}.
The theta functions guarantee that \eq{eqn:condition_dec} holds and that the NtNN distance, $y$, is larger than the distance to the NN, $x$. The effect of the clustering is encoded in the function $Q(x,y)$, the PDF to find the NN at $x$ and NtNN at $y$, which dimensionally scales as $Q\propto n^2$ and that includes the volume element factor $(4\pi \,x\,y)^2$. The factor $1/2$ accounts for the fact that each merger event involves a pair of PBHs. One can then rewrite
\begin{equation} 
{\rm d P} = \frac{1}{2}  Q(x,y)\, \theta(y-x)\, \theta(x_{\rm max}-x)\, \delta(y-\tilde{y}(t,x))\left|\frac{{\rm d} \tilde{y}}{{\rm d} t}(t,x)\right|\,{\rm d} x\, {\rm d} y,
\label{eqn:probability3BH_2}
\end{equation}
where the function $\tilde{y}(t,x)$ is obtained by inverting $t=t_{\rm m}(x,y)$.
The rate (per unit volume) at the time $t$ can then be obtained by direct integration, which thanks to the delta function reduces to:
\begin{equation}
\frac{dR}{dt} =n \int {\rm d P} =\frac{n}{2} \int_0^{x_{\rm max}} Q(x,\tilde{y}(x,t)) \mbox{ }\theta(\tilde{y}(x,t)-x)\left|\frac{{\rm d} \tilde{y}}{{\rm d} t}(t,x)\right|\,{\rm d}x\,.
\label{eqn:rate}
\end{equation}
Note that the above formalism---virtually used in all the literature on the subject---is only strictly valid for a monochromatic PBH mass function.

\section{Nearest-neighbour distributions}\label{NNdist}

Let us consider a statistically isotropic distribution of PBHs.  Given a PBH, we are interested in the probability of its nearest neighbour (NN)  and its next-to-nearest neighbour (NtNN) being located away from it between the distances $x$ and $x+{\rm d}x$ (for the NN) and $y$ and $y+{\rm d}y$ (for the NtNN), with $y>x$. This probability is of capital interest for the formation of binaries. 
Below we provide a justification for the expression of such a distribution $Q(x,y)$ (see eqs.~(\ref{eqn:probability3BH_1}) and (\ref{eqn:probability3BH_2})) in terms of  the two-point correlation function of PBHs, $\xi(r)$ (i.e.\ the excess probability over random),  see also \cite{Raidal:2017mfl}. A more formal and detailed discussion is provided in Appendix~\ref{derivation}.

Let us first assume that $\xi(r) =0$ for all $r>0$ --and the same for all higher order correlation functions-- and, therefore, the counting of PBHs is a  spatial homogeneous Poisson process. There are two key properties of such a point process. First, the expected number of points --PBHs in our case-- in a given region of space depends only on its volume; and second, the probabilities of finding any number of points in disjoint regions are independent of each other. Concretely,  the probability that $j$ PBHs occupy  a region  of comoving volume $V$ is, by assumption, 
\begin{align} \label{poss}
p(j,V)={n^jV^j}\exp(-nV)/{j!}\,,
\end{align}
and thus the {\it expected} number of PBHs in a volume $V$ is just $N=n\, V$. Clearly, the probability that there are no PBHs in a ball of radius $x$ and volume $V_x$ is simply given by $p(0,V_x)=\exp(-4\pi x^3 n/3)$. Using the assumption of independence of probabilities for disjoint regions,  the probability that any PBH has its nearest neighbour at a distance between $x$ and $x+\Delta x>x$ is given by the product $p(0,V_x)\,p(1,V_{\Delta x}-V_x)$, where $V_{\Delta x}-V_x= 4\pi((x+\Delta x)^3-x^3)/3$ is the volume of the spherical shell of radius $x$ and thickness $\Delta x$. In the limit of  $\Delta x\rightarrow 0$ we get that the (differential) probability of the NN being at a distance between $x$ and $x+dx$ is 
\begin{align}
P({\rm NN},x)dx=4\pi \,n \exp(-4\pi x^3 n/3)\, x^2 dx\,,\label{HertzP}
\end{align}
which is nothing but Hertz's formula \cite{Hertz1909}. Similarly, the probability of having no PBH in the volume contained in between $x$ and $y>x$ is $p(0,V_{y}-V_x)$; and the probability of having a PBH at a distance between $y$ and $y+\Delta y$ is just $ \,p(1,V_{\Delta y}-V_y)$. The joint (differential) probability distribution of having the NN at a distance between $x$ and $x+dx$ and the NtNN at a distance between $y$ and $y+dy$ (with all distances measured from a random PBH) is the product:
 $p(0,V_x)\,p(1,V_{\Delta x}-V_x)\, p(0,V_{y}-V_x)\, p(1,V_{\Delta y}-V_y)$,
in the limit ${\rm d} x\rightarrow 0$ and ${\rm d} y\rightarrow 0$, i.e.\
\begin{align} \label{joint}
P[({\rm NN},x)\,\&\,({\rm NtNN},y)]\, {\rm d} x\,{\rm d} y=Q(x,y){\rm d} x\,{\rm d} y= 16\pi^2\, n^2 \exp(-4\pi y^3 n/3)\,x^2\,y^2\,{\rm d} x\,{\rm d} y\,,\: y>x\,,
\end{align}
(which assumes $N-1\simeq N$).

These expressions can be heuristically extended to the case of a non-zero correlation function by writing eq. \eq{poss} as $p(j,V)={N^j}\exp(-N)/{j!}$  plus the replacement $N\to N=n \int_V (1+\xi)$. Then, the generalization of eq.~(\ref{HertzP}) writes
\begin{align}
P({\rm NN},x)dx=4\pi \,n \, (1+\xi(x))  \exp\left(-4\pi\, n \int_0^x \left(1+\xi(s)\right)s^2\,{\rm d} s \right)\, x^2 dx\,,\label{HertzGen}
\end{align}
and the function $Q$ of \eq{eqn:probability3BH_2} is
\begin{align}
Q(x,y)= 16\pi^2\,x^2\,y^2\,n^2 \exp\left[-4\pi\,n\int_0^y \left(1+\xi(s)\right)s^2\,{\rm d} s\right](1+\xi(x))\,(1+\xi(y))\,,\label{Qwithxi}
\end{align}
which we will use in the next section to compute the merger rate. Eq.~(\ref{Qwithxi}) matches the expression used in~\cite{Raidal:2017mfl}.
In Appendix~\ref{derivation}, a more formal derivation of these results is provided, showing that if the PBH spatial distribution is non-Gaussian, the above result
does not hold anymore.

\begin{figure}[h]
\begin{center}
\includegraphics[width= 0.65 \textwidth]{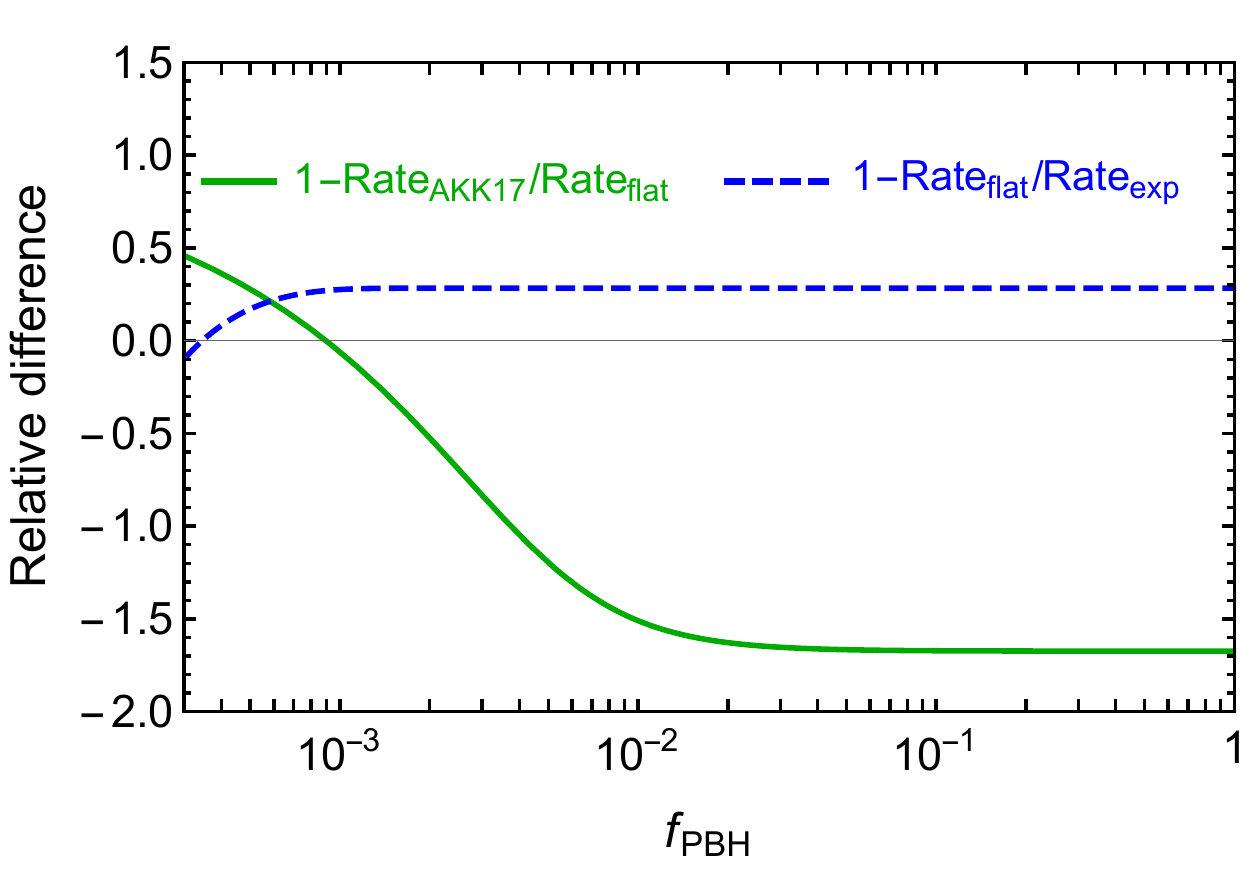}
\caption{Comparison of the merger rate obtained using different prescriptions for the nearest-neighbour distributions, eqs.~\eq{Qwithxi}, \eq{eq:flat}, and eq. (34) in~ref.~\cite{Ali-Haimoud:2017rtz}. We consider PBHs of a common mass $M=30\,M_{\odot}$ and we compute the merger rates at the present time ($t=13.7\,$Gyr), for an un-clustered case ($\xi$=0).} \label{excut} 
\end{center}
\end{figure}

It is worth noting that in some recent literature an approximation has been used: A flat spatial distribution with a sharp cut-off at $y=d\equiv\left(4\pi n/3\right)^{-1/3}$, see e.g.~\cite{Sasaki:2018dmp} 
\begin{equation}
Q_{\rm flat}(x,y)=(4\pi)^2 x^2\,y^2\, n^2 \Theta(y-x)\Theta(d-y)\,.
\label{eq:flat}
\end{equation}
We checked that for the physically most interesting range $f_{\rm PBH}\sim 10^{-3},$ eq.~(\ref{eq:flat}) leads to about a 30\% underestimate of the present merger rate  ($t=13.7\,$Gyr), 
as graphically reported in the blue dashed curve in fig.~\ref{excut}.

Within some minor analytical approximations and for the case $\xi(r)=0$, ref.~\cite{Ali-Haimoud:2017rtz} estimated the current merger rate including the effect of torques due to {\it all} neighbours.
The green solid curve of fig.~\ref{excut} compares the  present merger rate computed via eq.~(34) in ref.~\cite{Ali-Haimoud:2017rtz} to eq.~(\ref{eq:flat}). Accidentally, the curve is close to zero at values $f_{\rm PBH}\simeq 10^{-3}$, which are the most interesting ones corresponding to current upper bounds. However, at  values $f_{\rm PBH}\gg 10^{-3}$ the rate in~\cite{Ali-Haimoud:2017rtz} becomes up to two or three times higher than the rate computed according to eq.~(\ref{eq:flat}), suggesting that the corrections due to PBHs farther away acquire some importance for sufficiently dense PBH configurations.

\section{PBH abundance and correlation function}\label{correl}

In the previous section we derived an expression for the joint probability distribution function for the positions of the NN and the NtNN to any PBH, using the two-point correlation function,  $\xi(r)$, which is determined by both the PBH formation mechanism and the subsequent evolution under gravity.  As it is customarily done, we neglect the evolution of the system of PBHs between  PBH formation and  binary formation (i.e.\ we consider their clustering frozen after the PBH stop forming), and assume that the PBH formation mechanism is the collapse of large density fluctuations. To the best of our knowledge, this is the first time such a computation is explicitly performed, with the partial exception of an approximate and only parametric estimate presented in~\cite{Raidal:2017mfl}.

The basic idea of PBH formation in a cosmological context is that, if the energy density on some scale exceeds a critical threshold, a PBH forms when that scale becomes comparable to the Hubble radius, with a mass that is (a fraction $\gamma$ of) the mass within the Hubble horizon at that time. Numerical and analytical estimates of the efficiency of this process indicate that $\gamma\lesssim 1$ (see~\cite{Carr:2016drx} for a discussion) and, for simplicity,
we will adopt $\gamma=1$. The mass $M$ of a single PBH (when it forms) can be then parametrized in terms of the (comoving) scale $R$, which is the characteristic radius of the Hubble patch at formation time. 
In particular, for the illustrative calculations below, we assume the simplest and most widely used ``recipe'', namely that a PBH  forms when the energy density  fluctuation smoothed over the distance scale $R$, $\delta_R$, is above a threshold $\delta_c$. In general, $\delta_c$ depends on the equation of state of the medium and the shape of the fluctuation that triggers the collapse~\cite{Nakama:2013ica,Harada:2015yda}. For the sake of simplicity we assume $\delta_c=0.45$, which is commonly taken in the literature. Note that it has been recently argued \cite{Yoo:2018kvb,Germani:2018jgr} that this procedure is too simplistic (with possible order of magnitude errors in the mass and especially the abundance of the PBHs produced) and should be replaced by more local and less universal criteria as far as the dependence from the primordial power spectrum is concerned (to be calibrated on numerical studies of PBH formation). Since we are interested in estimating the {\it relative} impact of including the effect of $\xi(r)$ on the PBH binary formation, we will neglect these refinements, whose precise extent and implementation still needs to be clarified. However, we note that the study of these issues is of uttermost importance in assessing the viability of actual PBH production models, and the formalism below should be consider only as an approximate guidance to the actual phenomenon of the collapse.

Assuming the basic model of PBH formation, the two-point correlation function of PBHs, $\xi(r)$, is the excess probability  over random of two fluctuations exceeding
the threshold $\delta_c$ at a (comoving) distance $r$ of each other:
 \begin{equation}
\label{1pxi}
1+\xi(r)= \frac{p(\delta(0)>\delta_c, \delta(r)>\delta_c)}{P_1^2}=\frac{p(\delta(r)>\delta_c | \delta(0)>\delta_c)}{P_1}\,,
\end{equation}
where $P_1=p(\delta(0)>\delta_c)$ is the probability to exceed the threshold ``at a point'' (i.e. within a Hubble-sized patch, here considered pointlike), arbitrarily set as the origin of coordinates. We use $p(\delta(0)>\delta_c, \delta(r)>\delta_c)$ to denote the joint probability of finding a fluctuation above threshold at the origin and at a distance $r$ from it; whereas $p(\delta(r)>\delta_c | \delta(0)>\delta_c)$ is the conditional probability of finding a fluctuation above threshold at distance $r$, given an existing one (also) above threshold at the origin. In general, since any probability is smaller or equal than 1, we have $1+\xi(r)\leq P_1^{-1}\:\:\:\forall\, r$, as discussed in~\cite{Ali-Haimoud:2018dau}. There, it has been further
argued  that since $p(\delta(r)>\delta_c | \delta(0)>\delta_c)=1$ if $r= 0$, then $1+\xi(0)= P_1^{-1}$. However, one must be careful with the (interpretation of the) extrapolation to very small scales: in fact, $\xi(r)$ not only includes the correlation between a pair of distinct PBHs, $\xi_{\rm red}(r)$, but also the contribution to the autocorrelation of regions collapsing into a single PBH (so-called ``self-pairs''), eventually assuming the limiting (Poisson noise) value of $\delta(r)/n$ at zero distance (see~\cite{Desjacques:2018wuu} and refs. therein). $\xi_{\rm red}(r)$, the physically interesting ``reduced'' correlation function between distinct PBHs labelled by $R$, only coincides with $\xi(r)$ at scales $\gg R$, being equal  to $-1$ at smaller scales, since there is zero probability of finding two PBHs within a single patch . A toy-model for $\xi_{\rm red}(r)$ could thus be written as follows (see also~\cite{Desjacques:2018wuu})
\begin{equation}
\label{xred}\xi_{\rm red}(r)\simeq 
\begin{cases}
  -1 & \text{for $r<D$}\,,\\
    \xi(r)& \text{for $r\geq D\,,$}\,
  \end{cases}
  \end{equation}
  with the  naive expectation  $D\sim {\rm few}\:R$.

The smoothed density perturbation is most simply obtained in Fourier space, where it 
can be expressed as a convolution $\tilde\delta_R(k)=\tilde\delta(k) W_R(k)$ in terms of a conveniently chosen window function, $W_R(k)$.\footnote{To avoid clutter, in the following we use the same symbols for quantities and their Fourier transforms.}
Then, if the smoothed density contrast peaks at a comoving scale $k\equiv 1/R$, 
the mass-scale relation 
is
\begin{equation}
\label{eqn:Mk}
M(k)  \simeq 30\,q\, M_{\odot}  \left( \frac{k}{2.57  \mbox{ }10^6 \mbox{ Mpc}^{-1}} \right)^{-2}  \left(\frac{g_*}{50}\right)^{-1/6}\,,
\end{equation}
where $g_*$ is the number of the relativistic degrees of freedom at the time of formation, which in most models falls within the radiation domination epoch,
and $q\sim\mathcal{O}(0.1-1)$ depends on the choice of window function, e.g. $q=1$ corresponds to a top-hat window function in $k-$space, and intermediate and lower factors
correspond respectively to a Gaussian or a top-hat window function in real space.
The above relation follows from writing $M=V(R) (1+z_R)^{-3}  \rho(z_R)$, where $z_R$ is the redshift at which the scale $k=1/R$ enters the Hubble ``horizon'' and the over-density collapses, $\rho(z_f)$ is the background energy density at that time, and $V(R)$ is the comoving collapsing volume,
conventionally defined by the relation $W_R(0) V(R)=1$, which gives $V(R)=6 \pi^2 R^3$ for the top-hat window function in $k-$space.
Incidentally, 
adopting the prescription described above for $q=1$, we find a good agreement with the results of ~\cite{Yoo:2018kvb} for the resulting PBHs mass.

In order to proceed further, let us assume that the fluctuations $\delta$ of  the (radiation) background are Gaussian distributed, and described by a dimensionless spectrum  $\mathcal{P}_\delta(k)$.  Their smoothed variance over a scale $R$ is then
\begin{equation}
\label{eq:sigma}
\sigma_R^2\equiv \int_0^\infty \frac{{\rm d} k}{k}\mathcal{P}_\delta(k)W_R(k)^2\,,
\end{equation}
and the {\it cumulative} fraction of the energy density collapsing into PBHs of mass above $M$ is
\begin{equation}
\label{eq:beta}
\beta_R =\frac{1}{\sqrt{2\pi\sigma_R^2}} \int_{\delta_c}^\infty {\rm d}\delta\, \exp\left(-\frac{\delta^2}{2\sigma_R^2}\right)=\frac{1}{2}\mbox{Erfc}\left(\frac{\nu_R}{\sqrt{2}}\right)\,,
\end{equation}
where $\nu_R\equiv \delta_c/\sigma_R$.

Given $\mathcal{P}_{\delta}(k) $, the (radiation) density correlation function is:
\begin{equation}
\xi_\delta(r)= \int_0^\infty \frac{dk}{k} \mathcal{P}_{\delta}(k)  \frac{\sin (k r)}{kr}W_R(k)^2\,,
\label{eq:xi_rad}
\end{equation}
such that $\xi_\delta(0)=\sigma_R^2$. Although implicit in the expression above, $\xi_\delta$ depends also on redshift.
If we denote with $H=H(a)$ the Hubble parameter and $a$ the scale factor of the Universe at a given epoch, the radiation spectrum $\mathcal{P}_{\delta}$ is determined by the spectrum of primordial fluctuations $\mathcal{P}_{\zeta}$ (which in most models is seeded by inflation):
\begin{equation}
\mathcal{P}_{\delta}(k,a)\simeq\frac{16}{81}\mathcal{P}_{\zeta}(k)
\begin{cases}
   \left(\frac{k}{aH}\right)^4 & \text{for $k<a\,H$},\\
    1 & \text{for $k\geq a\,H$}.
  \end{cases}
  \label{eq:Pdelta}
\end{equation}
The factor $\propto k^4$ acts as an infrared cutoff in an equation like (\ref{eq:xi_rad}), suppressing the effect of super-Hubble modes. The primordial spectrum $\mathcal{P}_{\zeta}(k)$ needs to comply with current CMB and large scale structure (LSS) measurements, but can deviate from its power-law shape and be much enhanced at smaller scales, relevant for the formation of the PBHs and their clustering. In practice, since the modes probed by the CMB and LSS only contribute to $\xi(r)$ at a level $\lesssim 10^{-4}$, we will ignore their effect in the following, and only compute the contribution to $\xi(r)$ of the putative small-scale enhancement associated to PBH production.

We can compute (\ref{1pxi}) explicitly in the Gaussian case, following~\cite{Kaiser:1984sw}. 
 The {\it cumulative} correlation function\footnote{See e.g. \cite{Cooray:2002dia}  for explicit formulas for differential correlation function, in the formally analogous halo model.}  for PBHs of mass above $M$ is \begin{equation}
1+\xi(r)= \frac{1}{2 \pi \beta_R^2} \int_{\nu_R} d x  \int_{\nu_R} d y \frac{1}{\sqrt{1-w(r)^2}} \exp\left[-\frac{x^2+y^2-2\,x\,y\,w(r)}{2(1-w(r)^2)}\right]\,,
\label{gaussianP2}
\end{equation}
where, following standard notation, we define $w(r)\equiv \xi_\delta(r)/\sigma_R^2$. Notice that whereas  $\xi_\delta$ is given by $\mathcal{P}_\delta(k)$, $\xi$ is determined by the smoothed peaks of $\delta$ above a certain threshold. We identify it with the correlation function of PBHs, assuming it is a good proxy for it.

The last expression can be evaluated analytically in some limits. A compact formula which is exact in the limit $\nu_R^{-1}\rightarrow 0$ has been recently presented in~\cite{Ali-Haimoud:2018dau}: 
\begin{equation}
1+\xi(r)\simeq (1+w(r))\frac{\mbox{Erfc}\left(\sqrt{\frac{1-w(r)}{1+w(r)}}\,\nu_R/\sqrt{2}\right)}{\mbox{Erfc}\left(\nu_R/\sqrt{2}\right)}\,.\label{gaussianP2approx}
\end{equation}
This has the correct limiting behaviour at low and high $r$, and  leads to errors not larger than {\cal O(10\%)} even when $\nu_R\simeq 1$.
A commonly used approximation, known as Kaiser's bias~\cite{Kaiser:1984sw}, which holds under the conditions $w\ll 1/\nu_R^2\ll 1$, is
\begin{equation}
\xi(r) \simeq \xi_{\rm red}(r) \simeq \nu_R^2w(r)=\frac{\nu_R^2}{\sigma_R^2}\, \xi_{\delta}(r)\,.\label{Kaiser}
\end{equation}
The latter equation is particularly transparent to interpret. It is equivalent to the statement that when PBHs form, at sufficiently large scales --where $\xi_{\delta}(r)$ is small enough-- they are highly biased tracers of the underlying density field, with linear bias factor $\nu_R/\sigma_R\gg 1$. Away from this limit, they are still tracers of the density field, but the link is non-linear and thus more complicated.

\begin{figure}[]
\begin{center}
\includegraphics[width= 0.79 \textwidth]{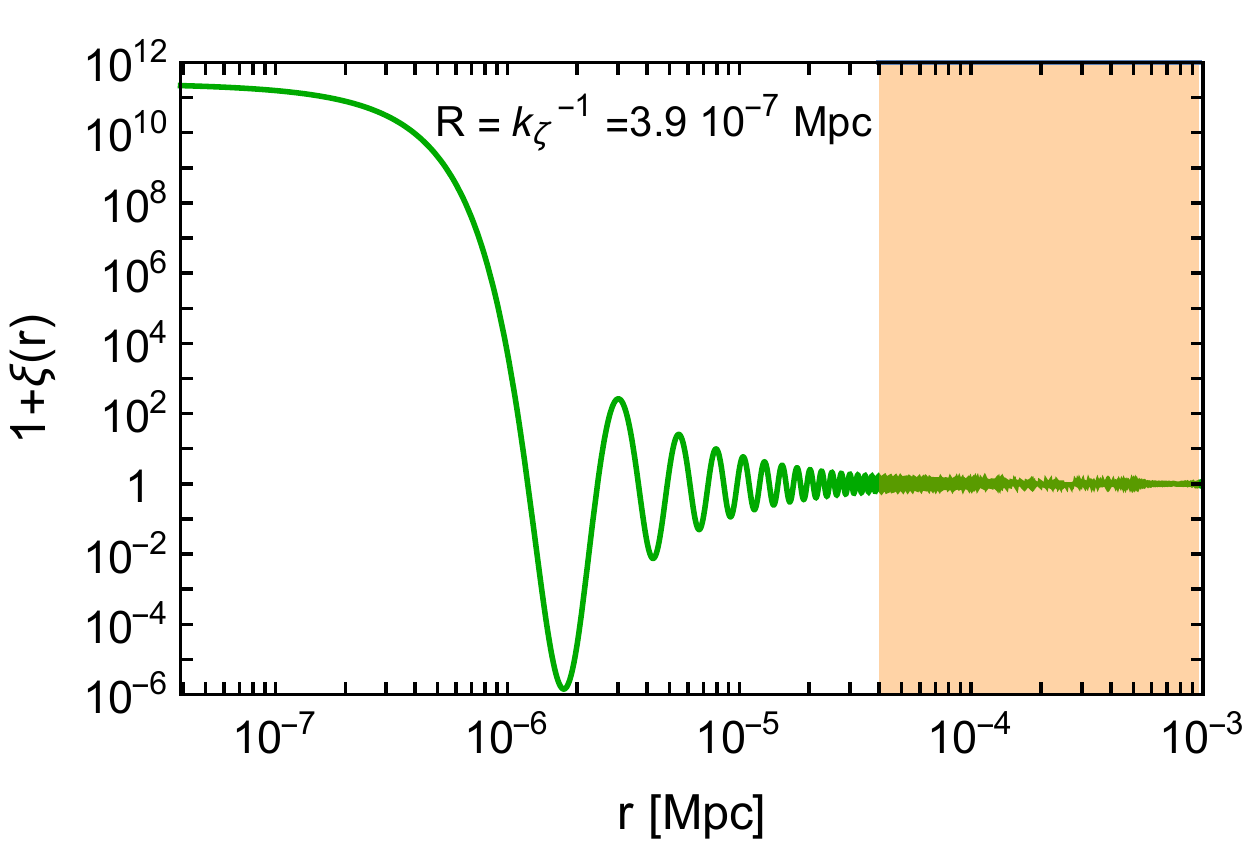}\vspace{1cm}
\includegraphics[width= 0.8\textwidth]{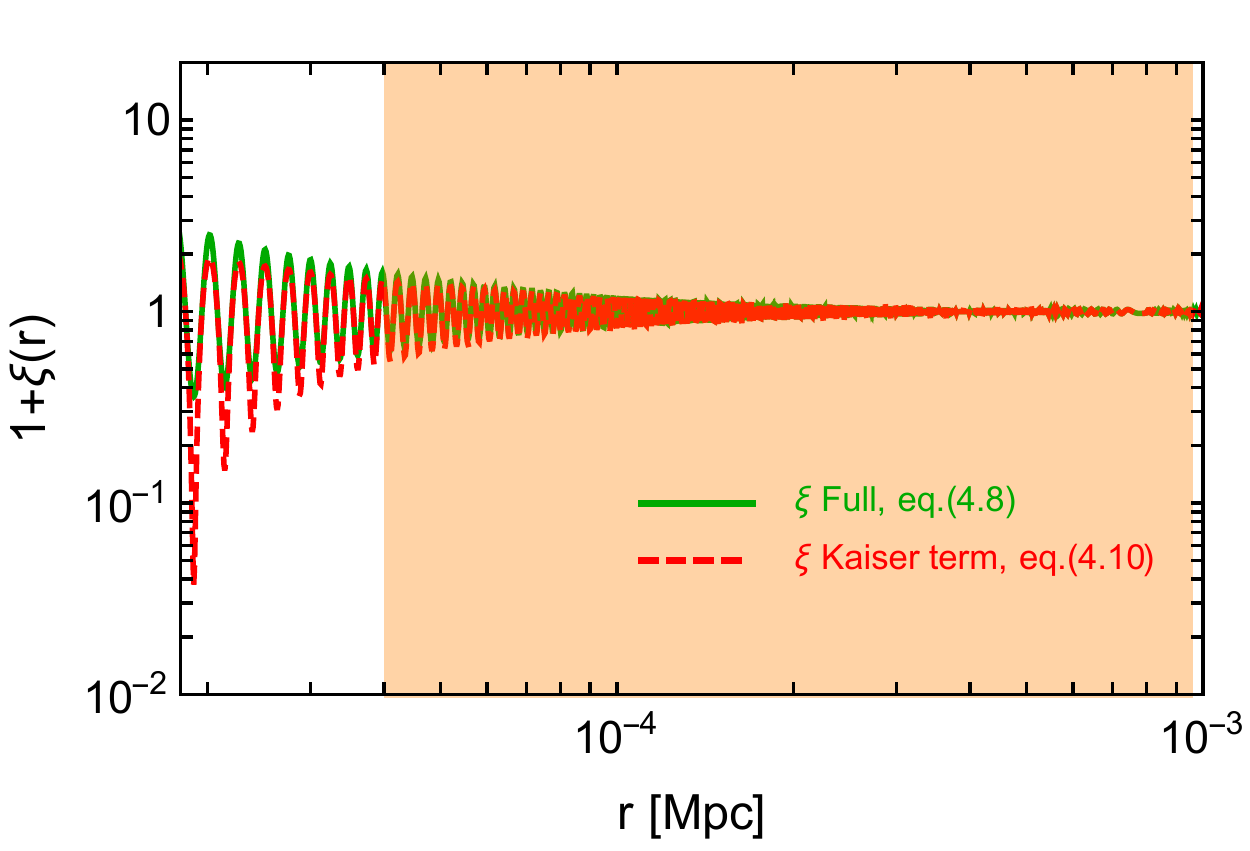}
\caption{The top panel shows the PBH two-point correlation function according to eq.~(\ref{gaussianP2}) for the power spectrum of eq.~(\ref{PSmono}). The shaded region denotes the approximate range of distances relevant for the calculation of the present merger rate ($t\simeq 14$ Gyr). The bottom panel compares eq.~(\ref{gaussianP2}) (shown also in the top panel) to the approximation of eq.~(\ref{Kaiser}) where the latter can be applied, zooming into the region of relevance for the merger rate.}	
\label{fig:xibh}
\end{center}
\end{figure}

As a concrete example,  let us consider the idealized case of a monochromatic primordial power spectrum, 
\begin{equation}
{P}_{\zeta}(k) = N_{\zeta} \delta(k-k_{\zeta}).\label{PSmono}
\end{equation}
This choice should be considered as representative of a narrow power spectrum above a smooth, power-law power spectrum extending to large scales (and adequately fitting the CMB and LSS data). 
Note that in this limit the function $w(r)$ is independent of $R$, as can be shown taking the ratio of \eq{eq:xi_rad} and \eq{eq:sigma}, and simply equal
to
\begin{equation}
w(r) =\frac{\sin (k_{\zeta}r)}{k_{\zeta}r}\,.
\end{equation}
If we additionally adopt
a top-hat window function in $k-$space, the limiting case of \eq{PSmono} also solves a technical complication, related to the cumulative nature of the mass function and the correlation function in their respective forms of eqs.~\eq{eq:beta} and \eq{gaussianP2}. Indeed, the PBH mass function becomes monochromatic, with a mass given by $M(k_\zeta)$ for $q=1$ in \eq{eqn:Mk}, and the cumulative 
nature of $\beta_R$ and $\xi(r)$ reduce to the functions themselves for this single mass value.~\footnote{Actually, if the gravitational collapse leading to PBH formation is a near-critical phenomenon (as approximately found already in~\cite{Niemeyer:1997mt}) even for a monochromatic primordial power spectrum, the resulting PBH mass function is expected to have a small (and negligible for our purposes) spread~\cite{Yokoyama:1998xd}.}
Note that for a Gaussian window function or a top-hat filter in real space, \eq{PSmono} would imply a broad mass distribution, with a sizable tail at low masses.\footnote{Drawbacks related to these alternative filters have been discussed in several contexts, see e.g.~\cite{Schneider:2013ria, Fairbairn:2017sil, Ando:2018qdb}.}
In the benchmark case that we adopt, the present  PBH abundance writes:
\begin{equation}\label{eq:beta3}
f_{\rm PBH}=\frac{\Omega_{\rm PBH}(M)}{\Omega_{\rm DM}} =   \frac{\beta_R(M)}{4\times 10^{-9}}  \left( \frac{g_*}{50} \right)^{-1/4} \left( \frac{M}{30\,M_{\odot}}\right)^{-1/2}\,.
\end{equation}
For definiteness, we choose the parameter $N_{\zeta}$ and $k_{\zeta}$ of the primordial power spectrum to obtain $M=30\,M_{\odot}$ and $f_{\rm PBH}=10^{-3}$, corresponding to $\nu_R=6.8$, in order to yield a predicted PBH merger rate similar to the one estimated by LIGO/Virgo on the basis of the GW events observed so far. 
Note that, as long as we neglect the (tiny) contribution of the large-scale CMB modes, the primordial power spectrum in eq.~(\ref{PSmono}) leads to the formation of PBHs only at the 
time when the scale $k_{\zeta}$ crosses the Hubble horizon. This manifests itself in a step-function behaviour of the quantities $\sigma_R$, $\beta_R$, 
 as it can be checked using eq.~(\ref{eq:sigma}) and eq.~(\ref{eq:beta}). Also notice how $\xi_\delta(r)$ and $\xi(r)$ are both contributed to by the single mode $k=k_\zeta$, thus are constant
 after it enters the horizon: the ratio of the PBH and the radiation power spectrum (at very small scales) or equivalently their respective correlation functions, is constant between the PBH
 formation and binary formation times.
This constant PBH two-point correlation function, referred to the case of eq.~(\ref{PSmono}), is shown in fig.~\ref{fig:xibh}. 
The numerical integration of \eq{gaussianP2} and the analytical approximation \eq{gaussianP2approx} lead to undistinguishable curves, on the scale of this figure.  At small distances, $\xi$ is dominated by the auto-correlation term (the analogous of the 1-halo term in the context of the halo model) responsible for the high-value plateau, while the correlation between different PBHs (or reduced correlation function $\xi_{\rm red}(r)$, corresponding to the ``2-halo term'' in the halo model analogy) takes over at larger distances, $r\gtrsim 2R$,  and is eventually
well described by eq.~(\ref{Kaiser}) (the dashed curve in the bottom panel), 
as expected. Since, as we shall show in a moment, the distances relevant for the formation of PBH binaries are much larger than $R$, we can assume for simplicity that  $\xi_{\rm red}(r)\simeq \xi(r)$ at the distances concerned. A more precise and detailed treatment, disentangling the two components (1-halo vs.\ 2-halo terms) contributing to the two-point correlation function, 
requires further physical input either from theory or, more likely, from simulations, than it is currently available.

As explained in sec.~\ref{basicf}, only PBHs separated by a distance smaller than $x_{\rm max}$ can form a binary system. In the example under consideration, $x_{\rm max}=9.6\times10^{-4}$ Mpc$\,\gg R$.  Moreover, there is also a minimum separation $x_{\rm min}$ for which the PBHs can form a binary and undergo merger rate at a time $t.$ This corresponds to configurations of PBHs leading to almost circular orbits, ${\sf e}=0.$ Focusing on the present merger rate ($t\simeq 14$ Gyr) we find $x_{\rm min}=4\times10^{-5}$ Mpc, again $\gg R$.
Notice that the probability for the formation of a binary, and thus the merger rate in \eq{eqn:rate},  is dominated by configurations of PBHs with the largest possible distances, simply because the volume of a spherical shell and therefore the probability to find a PBH increases, at least below the scale $d=\left(4\pi n/3\right)^{-1/3}$.

As shown in fig.~\ref{fig:xibh}, at the relevant distances the correlation function is $|\xi|\ll 1,$ therefore the PBH population can be considered to be distributed almost homogeneously.
A direct calculation of the merger rate including the PBH correlation function confirms this statement to sub-percent level. The same results hold for the stochastic GW background produced by the PBH mergers. In this case, one should integrate the GW signal produced by the merger of PBHs occurred at early times. We still find, in the example under consideration, that the clustering of PBHs is irrelevant up to very high redshift.
These conclusions are not surprising: Looking at 
\eq{gaussianP2} it is easy to realize that for a narrow primordial power spectrum, like the one in \eq{PSmono}, $\xi$ is suppressed at distances $r\gg k_{\zeta}^{-1}.$

\begin{figure}[t]
\begin{center} 
\includegraphics[width= 0.65 \textwidth]{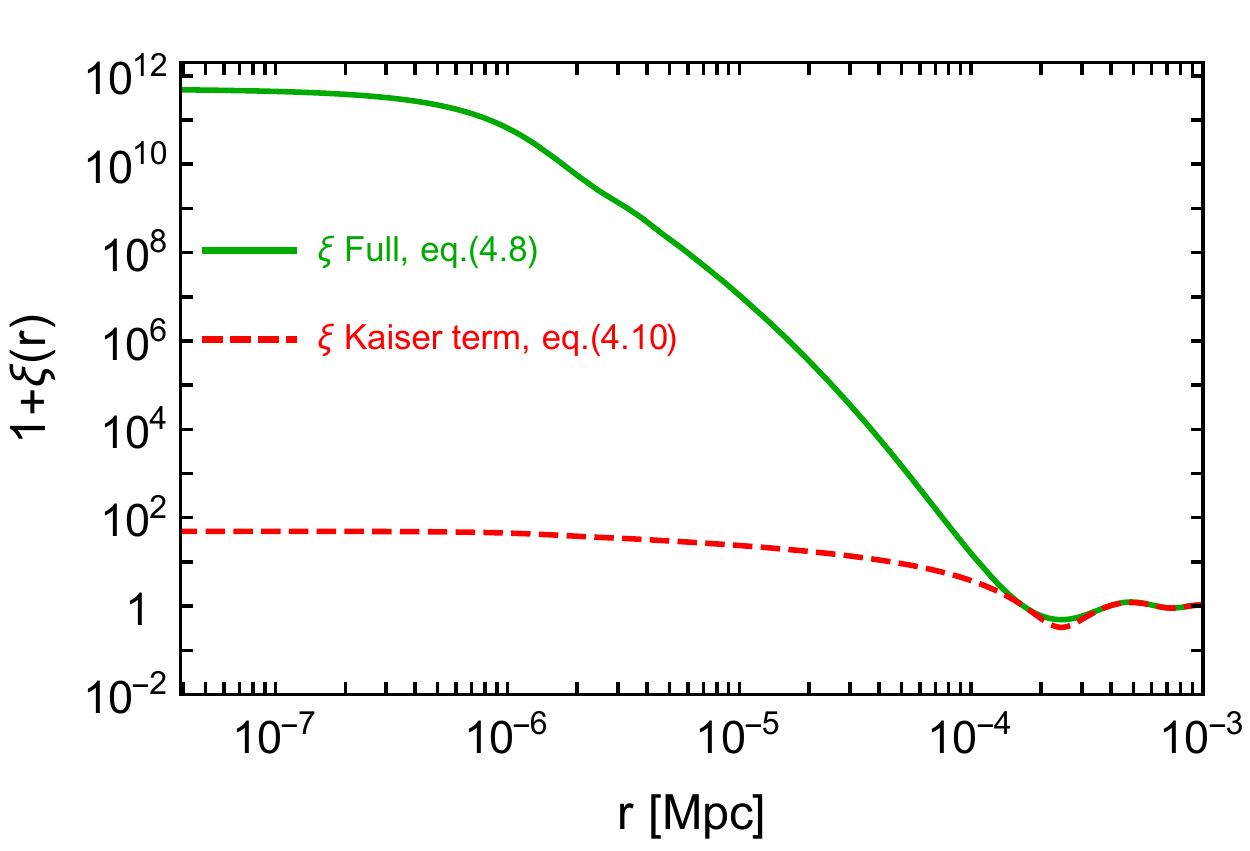}
\caption{The PBH two-point correlation function according to eq.~(\ref{gaussianP2}) and the approximation eq.~(\ref{Kaiser}), for the broad power spectrum described in the text.}
\label{fig:xibhBROAD}
\end{center}
\end{figure}

However, these results need not to hold true in general. For a very broad peak in the primordial power spectrum, between $k_\zeta$ and $k_{\rm min}\equiv k_\zeta/c$ (with $c\gg1$) the situation is more complex: 
First of all, the PBH mass function is unavoidably extended, and a proper extraction of the reduced (``2-halo'') PBH correlation function would be needed for  
a detailed computation. Second, the PBH formation phenomenon is more complex {\it in time}: as longer modes enter the horizon
a) PBHs of higher masses start forming, possibly  engulfing previously formed lighter PBHs; b) lighter PBHs  continue forming, the density field at relatively smaller scales
being possibly pushed above threshold for collapse by the contribution of the longer modes having now entered the horizon. 
Analytically, this translates into the properties that: i) values $\sigma_{R'}$ with $R'>R$ become sizable; ii) at fixed $R$, $\sigma_{R}$ continues growing until the longest sizable
mode $k_{\rm min}$ in $\mathcal{P}_{\zeta}(k)$ enters the horizon, after which $\sigma_R$ becomes constant. 
As far as we know, this extended PBH formation process has never been studied in detail, notably in its time dynamics. It is therefore unclear how much the simple formalism previously summarized can approximate the actual physics in this situation. We limit ourselves to describe the situation at the time the PBH 
formation is over, after which both $\xi(r)$ and $\xi_\delta(r)$ (again, neglecting the tiny CMB long modes) stay constant. 
To assess if the impact of PBH clustering is potentially important, we check whether $\xi_{\rm red}(r)$ attains values at least of ${\cal O}(1)$ at scales relevant for PBH binary formation.
One can check, using~\eq{Kaiser}, that this condition is satisfied for $\nu_R\approx 7$ if the peak extends by  about two orders of magnitude in $k-$space, but still well below the scales probed by CMB and LSS.\footnote{Constraints from $\mu$-distortion in CMB and pulsar timing array experiments could be important for such broad power spectra~\cite{Ando:2018qdb}, although these bounds significantly weaken in presence of non-Gaussianity~\cite{Nakama:2016gzw,Nakama:2016kfq,Nakama:2017xvq}.}
In fig.~\ref{fig:xibhBROAD}, we show  $\xi(r)$ computed using eq.~(\ref{gaussianP2approx}) (solid line) and eq.~(\ref{Kaiser}) (dashed line), for a box-shaped primordial power spectrum extending from $k_\zeta$ to $k_{\rm min}=k_{\zeta}/200$, $k_{\zeta}$ being associated to a PBH mass of $30\,M_{\odot}$. As one can appreciate, it is not evident how to separate the contribution of the ``1-halo'' and ``2-halo'' terms for $10^{-6}\lesssim r/\mbox{Mpc} \lesssim 10^{-4}$, but:
i) at sufficiently large distances, eq.~(\ref{gaussianP2approx}) and eq.~(\ref{Kaiser})  coincide, reassuring us that in this range we are looking at the $\xi_{\rm red}$ of interest.  ii) for distances relevant for the merger of PBHs of $\simeq 30\,M_{\odot}$ (as long as the computation performed for the monochromatic case is still representative in that respect), the correlation function is larger than unity. 
In table~\ref{tab:dataparameters} we show the distances $r_*$ above which the inequality $|\xi(r)|<1$ holds for a ``top-hat'' primordial power spectrum in the range $k_{\zeta}/c <k <k_{\zeta}$ for different choices of $c.$
Comparing these values with the shaded region in fig.~\ref{fig:xibh} we tentatively conclude that PBH clustering could have an important quantitative impact on the determination of the merger rate only for very broad primordial power spectra, extending at least two orders of magnitude in $k-$space. 
This conclusion makes sense, intuitively:  if the power-spectrum is sufficiently large at the scales corresponding almost to the (inverse) distances relevant for the formation for the binaries, so that PBHs almost as massive as the horizon size contained within these distances can also form, the clustering of the (light) PBH binaries is sizably affected.
Yet, we emphasize that this is a rough estimate, and for quantitative statements on the viability and phenomenology of broad mass function PBH models more detailed computations are needed. 

\begin{table}[]
 \small
  \centering
 \renewcommand\arraystretch{2.2}
\begin{tabular}{|c|c|c|c|c|c|c|}
\hline
 & $c=500$ & $c=200$ & $c=100$ & $c=50$ & $c=20$ & $c=10$ \\
\hline
\hline
 $|\xi(r>r_*)|<1$ & $3.2 \times 10^{-4}$ & $1.3 \times 10^{-4}$ & $6.8 \times 10^{-5}$  & $3.5 \times 10^{-5}$ & $2.9 \times 10^{-5}$ & $1.6 \times 10^{-5}$
\tabularnewline
\hline
\end{tabular} 
\caption{ Values of $r_*$ such that for $r>r_*$ the inequality $|\xi(r)|<1$ 
holds for a ``top-hat'' primordial power spectrum $P_{\zeta}$ that is taken to be constant for $k_{\zeta}/c <k <k_{\zeta}$ and vanishing outside this range. We choose a value of $k_{\zeta}$ corresponding to $M(k_{\zeta})=30 M_{\odot}$ in the monochromatic mass case. Note that at the values of $r>r_*$ reported in this table, eq.~(\ref{Kaiser}) provides an acceptable approximation,
so that $\xi(r)\simeq \xi_{\rm red}(r)$.}
\label{tab:dataparameters}
\end{table}

\section{Conclusions}\label{sec:conclusions}
In this article, we have analyzed a couple of key ingredients entering the standard computations of primordial black hole (PBH) binary formation in the early Universe, going back to the seminal paper~\cite{Nakamura:1997sm}. While refinement and clarification efforts were already initiated soon after that pioneering publication, see e.g.\ \cite{Ioka:1998nz}, the problem of computing the merger rate accurately has become quite timely in the light of the GW events detected by LIGO/Virgo. It is in fact crucial in deriving what at face value appears to be the most severe constraint to the idea that these events  may not only be due to PBHs, but actually
come from a population accounting for the majority or the totality of the still unidentified dark matter (DM). According to standard prescriptions for the calculation of the rate of PBH binaries, the required PBH abundance is at most two or three orders of magnitude below the required amount to account for all the DM.
This conclusion could mean a fatal blow to the aforementioned idea, since it comes from a self-consistency argument,
as opposed to the other numerous constraints which invoke different type of physics, and are subject to their own and quite varied uncertainties and systematics. 

Complementary to other recent efforts to assess the robustness of the merger estimate \cite{Ioka:1998nz,Ali-Haimoud:2017rtz}, we have discussed a couple of potentially important factors, namely the role played by: 1) the nearest neighbours distribution; 2) the details of PBH clustering, assuming a standard formation mechanism from the collapse of large density fluctuations. Our formalism thus allows to incorporate naturally the effect of the PBH clustering. 

Concerning the former, we have put the discussion on firmer ground, including both a simple heuristic derivation of the probability distribution function in Sec.~\ref{NNdist}, and a more rigorous one in Appendix~\ref{derivation}.  There, we correct some errors in the literature and also show e.g. that the relatively simple expressions  commonly used cannot hold for non-Gaussian  spatial distributions of PBHs. We finally perform a brief comparative assessment of the errors associated to different treatments found in the literature.

There have been surprisingly few studies in the literature on the generic expectations for (not to speak of consequences of) the PBH clustering. The most notable exception, ref.~\cite{Chisholm:2005vm}, 
concluded that the PBHs were ``born in clusters'', an effect so large that it would imply a significant impact on their merger rate, some consequences of which were worked out in~\cite{Chisholm:2011kn}. Recently, \cite{Ali-Haimoud:2018dau} showed that this erroneous conclusion originated from mishandled extrapolations of some analytical formulae (notably eq.~(21) in~\cite{Chisholm:2011kn}, applied also at small~$r$). On this point, our calculations agree with~\cite{Ali-Haimoud:2018dau}. On the other hand, the results of~\cite{Ali-Haimoud:2018dau} only refer to scales comparable to the one of the horizon collapsing into PBH, and therefore do not concern directly the PBH binary formation problem, which involves scales entering significantly later into the horizon. This observable depends upon the primordial power spectrum at $k-$values significantly smaller than the ones associated to  formation of PBHs of  mass {\cal O}(10)$M_\odot$. 
In Sec.~4 of this article we performed such a calculation step-by-step, in the limiting case of a monochromatic primordial power spectrum. This corresponds to a case often treated in the literature, leading to a narrow or monochromatic PBH mass function. 
Not surprisingly, since this model lacks power at the binary-relevant scales, one finds that the non-trivial clustering in the formation of PBH binaries can be neglected, at least if an accuracy not better than $0.1-1\%$ is needed. For PBH models whose mass function is narrow enough, which constitute a large fraction of those described in the literature, simplistic calculations of the binary merger rate are thus sufficiently precise. In that sense, our study confirms the robustness of this observable in these models. However, our results also indicate that for extended mass functions, associated to broad features in $k$-space (extending over two decades or more), the effect of the clustering is potentially relevant and should be evaluated on a model-by-model basis. 

Needless to say, a few important simplifications still exist in virtually all studies on the merger rate of PBH binaries, whose impact remains to be investigated. Let us mention a few: 
\begin{itemize}
\item {\bf Toy-model recipes vs.\ numerical results.}
An actual validation of analytical recipes for PBH formation against numerical studies, which may also reveal some surprises, is still lacking. In the last years, some studies have done preliminary steps to tackle these problems~\cite{Nakama:2013ica,Harada:2015yda,Yoo:2018kvb,Germani:2018jgr}, but the actual impact on phenomenological consequences (notably the PBH mass function) remains to be investigated.
\item {\bf PBH formation for broad mass function.}
The whole formalism summarized in Sec.\ 2 is only directly applicable for a monochromatic mass function. A generalization for broad mass functions (including the possibility of a non-trivial clustering) is required to reliably assess the compatibility of rates measured by LIGO/Virgo with actual models of PBH formation.
\item {\bf Non-linear evolution.}
Till now, the gravitational evolution of the system from the epoch of PBH formation until the epoch of binary formation has been neglected. In many respects, this is conservative, and one also expects it to be a satisfactory approximation for close to monochromatic PBH mass functions. Yet, for multi-scale problems characteristic of broad mass function models, which are also the ones for which the initial clustering effect is potentially most important, some non-trivial effects may exist, which (unfortunately) require a dedicated numerical simulation to be studied. 
\item  {\bf Gaussianity.} Another ``elephant in the room'' concerns the Gaussian approximation, done in almost all the studies.
An important issue would be to assess the level of non-Gaussianity actually implied in realistic models of PBH formation. This is likely to affect importantly both the PBH abundance and their clustering, as argued e.g. in~\cite{Bullock:1996at,Ivanov:1997ia,Yokoyama:1998pt,Saito:2008em,Byrnes:2012yx,Young:2015cyn,Kawasaki:2015ppx,Franciolini:2018vbk}.
\end{itemize}
Definitely, even if the possibility that PBHs of tens of solar masses could account for the majority of the DM appears nowadays much more unlikely than a couple of years ago, this provocative idea
has had (and will still have in the near future) a healthy effect in triggering studies clarifying fascinating cosmological aspects.  If they exist, PBHs could serve as a probe of the early Universe at distance scales much smaller than those accessible with standard cosmological observations. Therefore, obtaning a quantitative precise understanding of their theoretical merger rate remains an attractive question to explore.

\small

\subsubsection*{Acknowledgments}

We thank Y.\ Ali-Haimoud, R.\ Alonso, S.\ Camera, A.\ Carmona,  A.\ Casas,  V.\ Desjacques, M.\ Regis, A.\ Riotto and H.Veerma\"e for discussions. The work of GB is funded by a {\it Contrato de Atracción de Talento (Modalidad 1) de la Comunidad de Madrid} (Spain), with number 2017-T1/TIC-5520. It has also been supported by the {\it ERC Advanced Grant SPLE} under contract ERC-2012-ADG-20120216-32042, by {\it MINECO} (Spain) under contract FPA2016-78022-P and {\it Centro de Excelencia Severo Ochoa Program} (Spain) under grant SEV-2016-0597. MT acknowledge support from the research grant {\it TAsP} (Theoretical Astroparticle Physics) funded by the Istituto Nazionale di Fisica Nucleare (INFN).

\appendix
\section{Formal derivation of the NN and NtNN distributions}\label{derivation}
Let us illustrate in a more formal way under which hypothesis the relations reported in the main text, such as eq.~(\ref{HertzGen}) and eq.~(\ref{Qwithxi}), hold. In the physics of many-body systems (for some general ref., see e.g.~\cite{LiqText}), one finds specialized literature on the topic of the first neighbour distribution in non-ideal fluids~\cite{TorquatoPRA,TorquatoPA}, which may serve as a starting point and benchmark comparison for our application to PBH.
These calculations have also been extended to the second and higher order neighbour distributions in~\cite{Mazur92,Bhattacharjee2003}. These previously published higher order neighbour probabilities have a poor matching to numerically generated random distributions with $\xi(r)=0$, as we checked with Monte Carlo studies both in one and two dimensions~\footnote{We thank Hardi Veerm{\"a}e for an independent check of this point.}.
Below, we identify the mistake in~\cite{Mazur92} and derive the correct expression, eq.~(\ref{correctP2}), which to the best of our knowledge appears here for the first time.  The article~\cite{Bhattacharjee2003} provides a continuum formulation of~\cite{Mazur92} from the beginning, and rewrites the results of~\cite{Mazur92} into a recursive form, which is fully equivalent to~\cite{Mazur92}. Our correction thus applies to this paper as well.
For ease of comparison, we stick to the notation of~\cite{Mazur92}.

In general, one can define the $\nu$ particle density distribution, $\rho_\nu$, depending on the $3\nu$ coordinates of the $\nu$ particles,
such as that
\begin{equation}
\rho_\nu({\bf r}_1,\ldots {\bf r}_\nu){\rm d}^3{\bf r}_1\ldots {\rm d}^3{\bf r}_\nu
\end{equation}
gives the distribution function to simultaneously find one particle in the infinitesimal volume ${\rm d}^3{\bf r}_1$  around ${\bf r}_1$, a second particle in the infinitesimal volume ${\rm d}^3{\bf r}_2$  around ${\bf r}_2$, etc~\cite{LiqText}. The integral over all the arguments and the whole volume obeys the normalization
\begin{equation}
\int {\rm d}^3{\bf r}_1\ldots \int{\rm d}^3{\bf r}_\nu \rho_\nu({\bf r}_1,\ldots {\bf r}_\nu)=\frac{N!}{(N-\nu)!}\,.\label{normalization}
\end{equation}

It is also useful to introduce a discrete notation.  For instance
\begin{equation}
\rho_1({\bf r})=\sum_i^N\langle \delta({\bf r}-{\bf r}_i)\rangle
\end{equation}
where the brackets are ensemble averages over all configurations of the $N$ particles in the system, occupying a volume $V$ such that $n=N/V$.
For a homogeneous system, $\rho_1({\bf r})=n$, with its integral over volume giving the number $N$ of particles. In general, $\rho_\nu$ has dimensions
$[n]^\nu$.

Similarly,
\begin{equation}
\rho_2({\bf r},{\bf r}')=\sum_{j\neq i}^N\sum_i^N\langle \delta({\bf r}-{\bf r}_i)\delta({\bf r'}-{\bf r}_j)\rangle\,,
\end{equation}
with $\rho_2({\bf r},{\bf r}')\propto (1+\Xi({\bf r},{\bf r}'))$, $\Xi({\bf r},{\bf r}')$ being  the two-point correlation function.
In particular, by integrating $\rho_2$ over all (the five) coordinates other than the radial distance $r$
between the particle of reference and a particle under consideration, one can define the function $\rho_2^*(r)$,
\begin{equation}
\rho_2^*(r=|{\bf r}_2-{\bf r}_1|)=\int{\rm d}^3{\bf r}_1\int {\rm d}\Omega_{2-1}\rho_2({\bf r}_1,{\bf r}_1+{\bf r}_2-{\bf r}_1)\,.
\end{equation}
This is related to the pair correlation function $G(r)$ (and the ``integrated'' two-point correlation function $\xi(r)$, depending only on the relative distance), defined via 
\begin{equation}
G(r)\equiv\frac{1}{N}\sum_{j\neq i}^N\sum_{i}^N\left\langle \delta(r-r_{ij})\right\rangle=4\pi r^2 n(1+\xi(r))=\frac{\rho_2^*(r)}{N}\,,
\label{Gdef}
\end{equation}
$r_{ij}$ being the relative distance of the point $i$ from $j$. When integrating the above expression over the whole volume, one obtains the total number of neighbours to a given particle, $N-1$, since the integral of $\rho_2^*$ gives $N(N-1)$, see eq.~(\ref{normalization}).

Similarly to the case of the two-point correlation function, by integrating  $\rho_\nu$ over the extraneous coordinates, one can obtain in general the function of the $\nu-1$ radial distance variables  $\rho_\nu^*(r_{21},\ldots,r_{\nu-1})$, 
whose dimension is $[r]^{1-\nu}$. This has the advantage of making the following discussion less cluttered, although all the steps below could be rephrased in therms of the full $\rho_\nu$'s.
Alternatively, one may think of them as the one dimensional version of the whole problem (for one dimension, $\rho_\nu^*$ coincides to $\rho_\nu$ apart for the fixing of the reference point).

We can now introduce a partition of $G(r)$ into ``neighbourship distribution'', $P(1,r)$, $P(2,r)$, \ldots $P(N-1,r)$ such that $P(i,r)$ is the  probability that the $i-$th neighbour is found between $r$ and $r+{\rm d}r$. The following consistency sum rule holds:
\begin{equation}
G(r)=P(1,r)+P(2,r)+\ldots+P(N-1,r)\,.\label{sumrule}
\end{equation}
In particular, the explicit expression of the differential probability $P(1,r)$, also denoted in the main text as $P({\rm NN},r)$,  writes in the discrete case as
\begin{equation}
P(1,r)\equiv \frac{1}{N}\sum_{j\neq i}^N\sum_{i}^N\left\langle \delta(r-r_{ij})\prod_{k\neq i,j}^N \int_r^\infty{\rm d}r'\delta(r'-r_{ik})\right\rangle\label{P1def}\,.
\end{equation}
Note the product factor, which ensures that all the other particles are outside the radius $r$: For a given configuration of points, the product over $k$ makes sure that only one value of $j$ out of the $N-1$ is singled out (the nearest to $i$), annihilating all the others, so that the sum actually contains only $N$ terms, and the function is correctly normalized as a probability. 
 By using the identity
\begin{equation}
\int_r^\infty{\rm d}r'\delta(r'-r_{ik})=1-\int_0^r{\rm d}r'\delta(r'-r_{ik})\,,\label{oneminusint}
\end{equation}
we can replace in eq.~(\ref{P1def}) to obtain
\begin{eqnarray}
P(1,r)&&=\frac{1}{N}\sum_{j\neq i}^N\sum_{i}^N\left\langle \delta(r-r_{ij})\right\rangle-\frac{1}{N}\sum_{k\neq j\neq i}^N\sum_{j\neq i}^N\sum_{i}^N\left\langle \delta(r-r_{ij})\int_0^r{\rm d}r'\delta(r'-r_{ik})\right\rangle+\nonumber\\
&&+\frac{1}{N}\frac{1}{2!}\sum_{l\neq k\neq j\neq i}^N\sum_{k\neq j\neq i}^N\sum_{j\neq i}^N\sum_{i}^N\left\langle \delta(r-r_{ij})\int_0^r{\rm d}r'\delta(r'-r_{ik})\int_0^r{\rm d}r''\delta(r''-r_{il})\right\rangle-\ldots\nonumber\\
&&\equiv G(r)+\frac{1}{N}\sum_{\nu=3}^N\frac{(-1)^\nu}{(\nu-2)!}\int_0^r {\rm d}r'\int_0^r{\rm d}r''\ldots \,\rho_\nu^*(r,r',r'',\ldots)\label{P1exp}\,.
\end{eqnarray}
Note that we wrote each piece in brackets in terms of the same sums leading to the normalization coming from eq.~(\ref{normalization}),  in order to easily extract the $\rho_\nu^*$ factors, which obviously share the same normalization. The factorials~\footnote{Note that these factorials are missing in~\cite{Mazur92}. It must be a (repeated) typo, since the final result the author arrives at in the first part of his article is correct.} 
 at the denominator correct for the overcounting of terms implied from these multiple sums, and follow from the expansion of the product of binomials of the form~(\ref{oneminusint}). Since all other indices but the $i,\,j$ ones are involved, one obtains $(\nu-2)!$ Also note that the sum over $\nu$ extending up to $N$ is only formal. In practice, for all systems to which this formalism is applied one has $\rho_\nu^*(r,\ldots)\to 0$ at the relevant distances $r$ above some $\tilde\nu\ll N$.  In the case of PBH of interest here, for instance, if we look at scales corresponding to distances $\sf a$ relevant for binary formation whose merger time is of ${\cal O}$(14)Gyr, the maximum conceivable number of relevant neighbours for a primordial
 spectrum peaked at $k_\zeta$ is of the order $({\sf a} k_\zeta)^3$, since one cannot pack more than one PBH per Hubble horizon at formation (and the actual number is probably much smaller than that).  In microscopic systems such as liquids, short-distance intermolecular repulsion introduces a similar cutoff, as argued in~\cite{Mazur92}. Hence, a possibly large but actually finite number $\nu\ll N$ of terms contributes sizably to the series in eq.~(\ref{P1exp}).

 In the large $N$ limit, if the following condition holds for any finite $\nu\geq 3$~\footnote{This is known as {\it dilute limit} in the specialized literature on the statistical mechanics of liquids as the previously cited articles~\cite{Mazur92,Bhattacharjee2003}, and it is a condition that always holds at {\it low density} provided particles interact via an isotropic pair potential.}:
\begin{equation}
\rho_\nu^*(r,r',r'',\ldots)=\frac{\rho_2^*(r)}{N}\rho_{\nu-1}^*(r',r'',\ldots)\,,\label{dilute}
\end{equation}
then eq.~(\ref{P1exp}) can be rewritten by extracting the term $\nu=2$ out of the sum, collecting terms, and relabeling $\ell=\nu-1$. 
Eq.~(\ref{dilute}) clearly cannot hold for non-Gaussian spatial distributions, since it would otherwise allow one to express all {\it connected} distributions in terms of the  $\nu=2$ one.
One thus derives
\begin{equation}
P(1,r)=G(r)\left[1-\frac{1}{N}\sum_{\ell=2}^{N-1}\frac{(-1)^\ell}{(\ell-1)!}\int_0^r {\rm d}r'\int_0^r{\rm d}r''\ldots \rho_\ell^*(r',r'',\ldots)\right]\equiv G(r)\left[1-S(r)\right]\,.\label{PG}
\end{equation}

Using condition~(\ref{dilute}) repeatedly, one can prove that
\begin{equation}
\frac{1}{N}\int_0^r {\rm d}r'\int_0^r{\rm d}r''\ldots \rho_\ell^*(r',r'',\ldots)=h(r)^{\ell-1}\,,\label{intident}
\end{equation}
where $h(r)$ is the primitive of $G$, i.e. ${\rm d}h/{\rm d}r=G$. We can then prove two identities:
\begin{itemize}
\item By taking the derivative of the function $S(r)$ defined in eq.~(\ref{PG}) and using eq.~(\ref{intident}) we arrive (in the large $N$ limit) at
\begin{equation}
\frac{{\rm d}S}{{\rm d}r}=\sum_{\ell=2}^{N-1}(-1)^\ell \frac{h(r)^{\ell-2}}{(\ell-2)!}G(r)=P(1,r)\,,\label{dSdr}
\end{equation}
hence
\begin{equation}
P(1,r)=G(r)\left[1-\int_0^rP(1,x)\,{\rm d}x\right] \,,\label{Hertzcond}
\end{equation}
\item Also,  using eq.~(\ref{intident}) we may direct sum the series in eq.~(\ref{PG}), yielding 
\begin{equation}
P(1,r)=G(r)\exp\left[-h(r)\right]=G(r)\exp\left[-\int_0^r {\rm d}x\,G(x)\right]\,.\label{GenHertz}
\end{equation}
\end{itemize}
Eq.~(\ref{GenHertz}) is nothing but the generalization of Hertz solution~\cite{Hertz1909} for a non-trivial correlation $G(r)$, and agrees with the expressions found by an independent method in~\cite{TorquatoPRA,TorquatoPA} as well as our  eq.~(\ref{HertzGen}) derived heuristically in the main text.
Apart for corrections of typos in several formulae (notably his eq.s (2.4), (3.1), and (3.2)),  the steps above have the same start and end-point than the derivation presented in~\cite{Mazur92}, and also correspond to the final result for $P(1,r$) in~\cite{Bhattacharjee2003}.

When moving to $P(2,r')$ (also denoted with $P({\rm NtNN},r')$ in the main text), the differential probability of the second neighbour being between $r'$ and $r'+{\rm d}r'$, ref.~\cite{Mazur92} actually argues {\it by analogy}:
If we subtract $P(1,r)$ from $G(r)$, it leaves the pair distribution of all neighbours farther than the first, and $P(2,r)$ then obeys an equation analogous to eq.~(\ref{GenHertz}),
apart for the replacement $G(r)\to G(r)-P(1,r)$. Loosely speaking, subtracting from $G(r)$ the differential distribution for the NN, the NtNN behaves as the new NN with respect to the subtracted distribution $G(r)-P(1,r)$ i.e.
\begin{equation}
P(2,r)=(G(r)-P(1,r))\exp\left[-\int_0^r {\rm d}x(G(x)-P(1,x))\right]\,.\label{GenHertz2}
\end{equation}
We find however that this heuristic argument is incorrect, and a different relation holds. Let us  start from the definition of
$P(2,r')$, the differential probability of the second neighbour being between $r'$ and $r'+{\rm d}r'$,
\begin{equation}
P(2,r')\equiv \frac{1}{N}\sum_{k\neq j\neq i}^N\sum_{j\neq i}^N\sum_{i}^N\left\langle \left(\int_0^{r'}{\rm d r}\delta(r-r_{ij})\right)\delta(r'-r_{ik})\prod_{l\neq i,j,k}^N \int_{r'}^\infty{\rm d}r''\delta(r''-r_{il})\right\rangle\,.
\label{P2def}
\end{equation}
An expansion analogous to eq.~(\ref{P1exp}) leads to
 \begin{eqnarray}
P(2,r')&&=\frac{1}{N}\sum_{k\neq j\neq i}^N\sum_{j\neq i}^N\sum_{i}^N\left\langle \left(\int_0^{r'}{\rm d r}\delta(r-r_{ij})\right)\delta(r'-r_{ik})\right\rangle\,-\nonumber\\
&&-\frac{1}{N}\frac{1}{1!}\sum_{l\neq k\neq j\neq i}^N\sum_{k\neq j\neq i}^N\sum_{j\neq i}^N\sum_{i}^N\left\langle \left(\int_0^{r'}{\rm d}r\delta(r-r_{ij})\right)\delta(r'-r_{ik})\int_0^{r'}{\rm d}r''\delta(r''-r_{il})\right\rangle-\ldots\nonumber\\
&&\equiv -\frac{1}{N}\sum_{\nu=3}^N\frac{(-1)^\nu}{(\nu-3)!}\int_0^{r'} {\rm d}r\int_0^{r'}{\rm d}r''\ldots \rho_\nu^*(r,r',r'',\ldots)\,,\label{P2exp}
\end{eqnarray}
where, as argued in eq.~(\ref{P1exp}), only the indices following from the expansion of the product of binomials have the further factorial at the denominator.

By using the condition~(\ref{dilute}),  eq.~(\ref{P2exp}) rewrites (at the second step, relabeling $\ell=\nu-1$)
 \begin{eqnarray}
P(2,r')&&=\left(\int_0^{r'}{\rm d}r\,G(r)\right) \left[G(r')-\frac{1}{N}\sum_{\nu=4}^N\frac{(-1)^\nu}{(\nu-3)!}\int_0^{r'} {\rm d}r''\ldots   \rho_{\nu-1}^*(r',r'',\ldots)\right]\nonumber\\
&&= \left(\int_0^{r'}{\rm d}r\,G(r)\right) \left[G(r')+\frac{1}{N}\sum_{\ell=3}^{N-1}\frac{(-1)^\ell}{(\ell-2)!} \int_0^{r'}{\rm d}r''\ldots  \rho_{\ell}^*(r',r'',\ldots)\right]\,.
\end{eqnarray}
In the large $N$ limit, a direct comparison with eq.~(\ref{P1exp}) shows that the sum over $\ell$ at the last line is nothing but $P(1,r')-G(r')$, hence deriving 
\begin{equation}
P(2,r)= P(1,r)\int_0^{r}{\rm d}x\,G(x)\,.\label{correctP2}
\end{equation}
The quantity we are interested to in the main text is $Q(x,y)$, such that $P(2,y)=\int_0^y Q(x,y){\rm d} x$. It is immediate to check that eq.~(\ref{correctP2}) is consistent with eq.~(\ref{Qwithxi})
while eq.~(\ref{GenHertz2}) reported in~\cite{Mazur92,Bhattacharjee2003} is not.  The previous derivation can be obviously generalized to higher neighbourship distributions. As another consistency check, one can quickly verify that in the one dimensional case with $\xi(r)=0$ the relation~(\ref{sumrule}) is satisfied in the large-$N$ limit. By direct calculation, one finds in fact
\begin{equation}
P(1,r)=n\,e^{-nr}\,,\: P(2,r)= n\,e^{-n\,r} (n\,r)\,,\:\ldots P(i,r)= n\,e^{-n\,r} \frac{(n\,r)^{i-1}}{(i-1)!}\,,
\end{equation}
from which it follows
\begin{equation}
n=G(r)=\lim_{N\to \infty}\sum_{i=1}^{N-1}n\,e^{-n\,r} \frac{(n\,r)^{i-1}}{(i-1)!}=n\,e^{-n\,r}e^{n\,r}\,.
\end{equation}
Note how the essential hypothesis for these relations to hold boils down to eq.~(\ref{dilute}), which also implies that in presence of non-Gaussianities the formalism requires substantial
revision.


\end{document}